\documentclass[
 amsmath,amssymb,
 aps,
 prb,
 floatfix,
 10pt,
 twocolumn]
{revtex4-2}

\usepackage{graphicx} 
\usepackage{dcolumn}
\usepackage{color}
\usepackage{bm}

\usepackage{subcaption}
\usepackage{multirow}

\usepackage{xcolor}

\begin{document}

\title{Actively trained magnetic moment tensor potentials for mechanical, dynamical, and thermal properties of paramagnetic CrN}
\author{Alexey S. Kotykhov}
\affiliation{Skolkovo Institute of Science and Technology, Skolkovo Innovation Center, Bolshoy boulevard 30, Moscow, 121205, Russian Federation}
\affiliation{Moscow Institute of Physics and Technology, 9 Institutskiy per., Dolgoprudny, Moscow Region, 141701, Russian Federation}
\author{Max Hodapp}
\affiliation{Materials Center Leoben Forschung GmbH (MCL), Leoben, Austria}
\author{Christian Tantardini}
\affiliation{CNR - Istituto Officina dei Materiali (IOM) Cagliari, Cittadella Universitaria, Monserrato (CA), 09042, Italy}
\affiliation{Department of Materials Science and NanoEngineering, Rice University, Houston, Texas 77005, United States of America}
\affiliation{Institute of Solid State Chemistry and Mechanochemistry SB RAS, ul. Kutateladze 18, 630128, Novosibirsk, Russian Federation}
\author{Konstantin Kravtsov}
\affiliation{Moscow Institute of Physics and Technology, 9 Institutskiy per., Dolgoprudny, Moscow Region, 141701, Russian Federation}
\author{Ivan Kruglov}
\affiliation{Moscow Institute of Physics and Technology, 9 Institutskiy per., Dolgoprudny, Moscow Region, 141701, Russian Federation}
\author{Alexander V. Shapeev}
\affiliation{Skolkovo Institute of Science and Technology, Skolkovo Innovation Center, Bolshoy boulevard 30, Moscow, 121205, Russian Federation}
\author{Ivan S. Novikov}
\affiliation{Skolkovo Institute of Science and Technology, Skolkovo Innovation Center, Bolshoy boulevard 30, Moscow, 121205, Russian Federation}
\affiliation{Moscow Institute of Physics and Technology, 9 Institutskiy per., Dolgoprudny, Moscow Region, 141701, Russian Federation}
\affiliation{Emanuel Institute of Biochemical Physics of the Russian Academy of Sciences, 4 Kosygin Street, Moscow, 119334, Russian Federation}

\begin{abstract}
    We present a protocol for automated fitting of magnetic Moment Tensor Potential explicitly including magnetic moments in its functional form. For the fitting of this potential we use energies, forces, stresses, and magnetic forces (negative derivatives of energies with respect to magnetic moments) of configurations selected with an active learning algorithm. These selected configurations are computed using constrained density functional theory, which enables calculating energies and their derivatives for both equilibrium and non-equilibrium (excited) magnetic states. We test our protocol on the system of B1-CrN and demonstrate that the automatically trained magnetic Moment Tensor Potential reproduces mechanical, dynamical, and thermal properties, of B1-CrN in the paramagnetic state with respect to density functional theory and experiments.
\end{abstract}

\maketitle

\section{Introduction}

Magnetic materials have a wide range of applications \cite{heck2013magnetic, krishnan2016fundamentals}, particularly in such areas as data storage, magneto-optics, medical diagnostics, etc.
In addition, upon heating, materials may become paramagnetic, requiring a model for paramagnetism in order to accurately predict material properties.
A number of studies \cite{crisan2002magnetochemical, smirnov2005importance, ekholm2010influence} have shown that modeling a material in the paramagnetic state as a non-magnetic system can lead to wrong results.
Therefore, it is necessary to explicitly consider magnetic disorder with non-zero local magnetic moments. 

Machine-learning interatomic potentials (MLIPs) have become a reliable tool in computational materials science.
Since 2020, many MLIPs explicitly including magnetic moments have been developed and used for investigating materials in the paramagnetic state.
In their paper \cite{eckhoff2021_mhdnnp}, the authors developed a high-dimensional neural network potential including magnetic moments in its functional form and used it to predict the Neel temperature in the Mn-O system.
In \cite{Novikov2022-mMTP}, it was shown that the magnetic Moment Tensor Potential (mMTP) reproduces the phonon spectra of Fe in the paramagnetic and ferromagnetic states.
In the work \cite{drautz2024_noncolACE}, the authors tested another magnetic MLIP, namely, non-collinear magnetic atomic cluster expansion potentials on the Fe system.
They predicted many properties of Fe including the Curie temperature.
In \cite{yu2024_spinGNN, yu2024_spinGNN++}, the authors developed spin-dependent graph neural network potentials and applied them to the BiFeO$_3$, CrI$_3$, and CrTe$_2$ systems. 
The Neel temperature was calculated for BiFeO$_3$ and the Curie temperature was computed for CrI$_3$ and CrTe$_2$.
More details on the above works can be found in the review paper on existing magnetic MLIPs \cite{Kostiuchenko2024_magneticMLIPs}.

In all of the above works, the training sets were constructed manually.
Here, we describe a protocol for automated mMTP fitting and automated construction of the training set.
This protocol includes active learning initially proposed for non-magnetic MTP in \cite{podryabinkin2017-AL} and constrained density functional theory (cDFT) calculations \cite{Gonze_2022,Tantardinin2025_Y114} with hard constraints on magnetic moments.
The proposed methodology enables for selecting configurations for the training set during any atomistic simulation, e.g., molecular dynamics or geometry optimization (relaxation). 
Constrained DFT enables us calculating the selected configurations with non-equilibrium (excited) magnetic moments.
We test the developed methodology on the chromium nitride (CrN) system in the B1 phase (rock-salt structure) in the paramagnetic state. Hereinafter, we will refer to this system as B1-CrN. We selected this system due to the extensive availability of experimental and theoretical results \cite{chase1998nist, zhang2010crn, alling2010effect, zhou2014structural}.
For describing the paramagnetic state, we average over different randomly disordered collinear magnetic states.
We show that, with such an averaging, we are able to reproduce the properties of B1-CrN in the paramagnetic state. 
Namely, we demonstrate that the fitted mMTP yields elastic constants, phonon spectrum, thermal lattice expansion coefficient, and specific heat capacity of this material, in agreement with DFT and experiments.

\section{Methodology} \label{Methods}

\subsection{Magnetic Moment Tensor Potential}

We used multi-component magnetic Moment Tensor Potential (mMTP) as an interatomic interaction model. This model was originally developed in \cite{Shapeev2016-mtp,gubaev2018machine} for non-magnetic materials and further extended to magnetic materials \cite{Novikov2022-mMTP,Kotykhov2023-cDFT-mMTP}.
We emphasize that mMTP includes only collinear magnetic moments in its functional form.

Before explaining mMTPs, we define an atomic configuration with periodic boundary conditions that approximates an infinite atomic system and a local atomic environment.
Let ${\bm x} = \{(\bm{l}_1, \bm{l}_2, \bm{l}_3); ({\bm r}_i,z_i,m_i), ~i=1,\ldots,N \}$ be a configuration with $N$ atoms in a supercell, each atom being encoded by its position ${\bm r}_i$, atomic type $z_i$, and collinear magnetic moment $m_i$. To approximate an infinite system, three lattice vectors $\bm{l}_1, \bm{l}_2, \bm{l}_3$ are also introduced for each configuration.
These vectors enable the replicating for atoms from a supercell. The position of the $i$-th replicated atom is ${\bm r}_i + n_1 \bm{l}_1 + n_2 \bm{l}_2 + n_3 \bm{l}_3$ (here $n_1$, $n_2$, and $n_3$ are integer numbers) whereas the atomic type $z_i$ and the magnetic moment $m_i$ are the same as in a supercell.
We denote a set of lattice vectors by $L = (\bm{l}_1, \bm{l}_2, \bm{l}_3)$, a set of positions in a supercell by $R = ({\bm r}_1, \ldots, {\bm r}_N)$, a set of atomic types in a supercell by $Z = (z_1, \ldots, z_N)$, and a set of scalar magnetic moments in a supercell by $M=(m_1, \ldots, m_N)$.

We further construct local atomic environments (or neighborhoods) for each $i$-th atom in a supercell.
To that end, we introduce a cutoff radius $R_{\rm cut}$, i.e. the radius vector of the $i$-th central atom of length $R_{\rm cut}$ and include in the neighborhood $\mathfrak{n}_i$ each $j$-th atom with $|{\bm r}_{ij}| = |{\bm r}_j - {\bm r}_i| \leq R_{\rm cut}$, the atomic types of the central atom $z_i$ and neighboring atoms $z_j$, as well as magnetic moments of the central atom $m_i$ and neighboring atoms $m_j$.

Having the configurations and local atomic neighborhoods well-defined, we now introduce mMTPs which can be written as a sum of per-atom contributions $V$:
\begin{equation}
\label{eq:energy}
 E = \sum\limits_{i=1}^N V (\mathfrak{n}_i) = \sum\limits_{i=1}^N \sum_{\alpha} \xi_{\alpha} B_{\alpha}(\mathfrak{n}_i),    
\end{equation}
where $\xi_{\alpha}$ are the parameters to be fitted and $B_{\alpha}$ are the basis functions. The basis functions are scalar contractions of the Moment Tensor Descriptors:
\begin{equation}
\label{eq:descriptors}
M_{\mu,\nu}(\mathfrak{n}_i)=\sum_{j} f_{\mu}(| {\bm r}_{ij}|,z_i,z_j,m_i,m_j) {\bm r}_{ij}^{\otimes \nu},    
\end{equation}
where $f_{\mu}(| {\bm r}_{ij}|,z_i,z_j,m_i,m_j)$ is a radial part describing pairwise interactions between a couple of atoms \textit{i} and \textit{j}, $\mu$ is the number of radial functions, and ${\bm r}_{ij}^{\otimes \nu}$ is an angular part describing many-body interactions; the outer product ``$\otimes$'' of the vector ${\bm r}_{ij}$ is performed $\nu$ times.
The radial part has the following form:
\begin{equation} \label{eq:radial}
\begin{array}{c}
\displaystyle
    f_{\mu}(| {\bm r}_{ij}|,z_i,z_j,m_i,m_j) = \sum_{\zeta=1}^{N_{\phi}} \sum_{\beta=1}^{N_{\psi}}\sum_{\gamma=1}^{N_{\psi}}c_{\mu,z_i,z_j}^{\zeta,\beta,\gamma} \times
\\
\displaystyle
    \phi_{\zeta}(|\bm r_{ij}|) \psi_{\beta}(m_i)\psi_{\gamma}(m_j) (R_{\rm cut} - |\bm r_{ij}|)^2,
\end{array}
\end{equation}
where $c_{\mu,z_i,z_j}^{\zeta,\beta,\gamma}$ are the parameters to be fitted, $\phi_{\zeta}(|\bm r_{ij}|)$, $\psi_{\beta}(m_i)$, and $\psi_{\gamma}(m_j)$ are Chebyshev polynomials of the orders $\zeta$, $\beta$, and $\gamma$, respectively, $N_{\phi}$ and $N_{\psi}$ are the numbers of these polynomials.
The argument $|\bm r_{ij}|$ of the polynomial $\phi_{\zeta}$ is defined on the interval $(R_{\rm min},R_{\rm cut})$ where $R_{\rm min}$ is the minimal distance between a couple of atoms \textit{i} and \textit{j} in the training set. 
The arguments $m_i$ and $m_j$ of the polynomials $\psi_{\beta}$ and $\psi_{\gamma}$ are defined on the interval $(-M_{\rm max}^{z_i},M_{\rm max}^{z_i})$. The value $M_{\rm max}^{z_i}$ depends on the atomic type $z_i$ and its physical meaning is the maximum magnetic moment for an atom of type $z_i$ in the training set.

\begin{figure*}[!ht]
\centering
\includegraphics[scale=0.45]{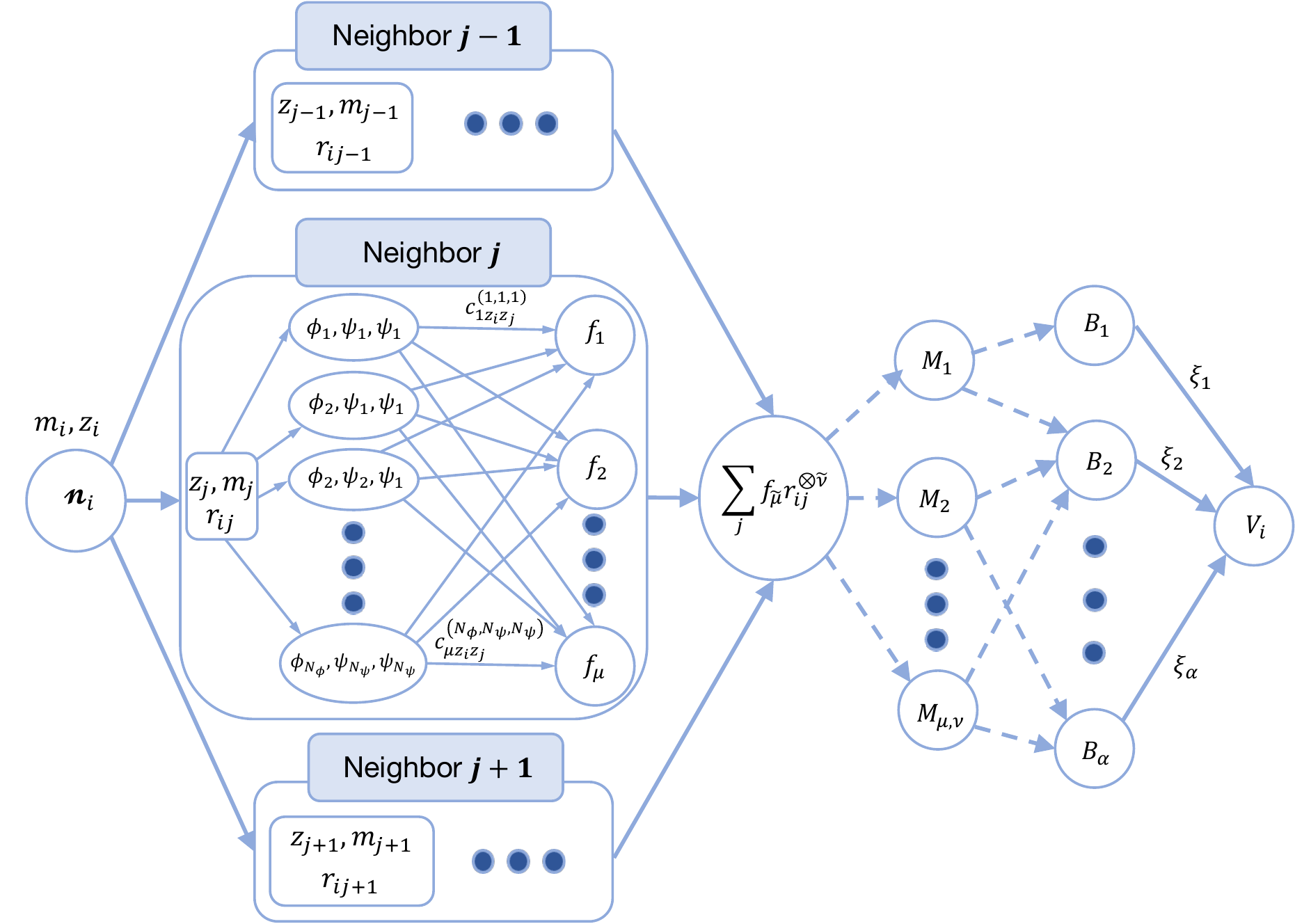}
\caption{Computation scheme of the per-atom energy contribution $V_i$ in magnetic MTP.}
\label{fig:mMTP}
\end{figure*}

We compute the basis functions $B_{\alpha}$ from all possible contractions of the Moment Tensor Descriptors $M_{\mu, \nu}$, i.e., the tensors of rank $\nu$, yielding a scalar. To restrict the number of the basis functions $B_{\alpha}$ in the mMTP functional form \eqref{eq:energy}, we introduce the level of Moment Tensor Descriptors ${\rm lev} M_{\mu, \nu} = 2 + 4 \mu + \nu$. If the descriptors $M_{\mu_1, \nu_1}$, $M_{\mu_2, \nu_2}$, \ldots, are included in the contractions yielding $B_{\alpha}$, then ${\rm lev} B_{\alpha}$ = $(2 + 4\mu_1 + \nu_1)$ + $(2 + 4\mu_2 + \nu_2)$ + $\dots$. We further introduce the number ${\rm lev_{\rm mMTP}}$, i.e., the level of mMTP and include in \eqref{eq:energy} only the basis functions with ${\rm lev} B_{\alpha} \leq {\rm lev_{\rm mMTP}}$. The described scheme for $V_i$ calculating is shown in Fig. \ref{fig:mMTP}.

We denote a set of the mMTP parameters by $\bm \theta = \{\xi_{\alpha}, c_{\mu, z_i, z_j}^{\zeta, \beta, \gamma} \}$ and the energy \eqref{eq:energy} by $E = E({\bm \theta}, R, Z, L, M)$. We denote the mMTP energy by:
\begin{equation}
\label{eq:mMTPenergy}
E^{\rm mMTP}(\bm \theta) = \dfrac{E(\bm \theta, R, Z, L, M)+E(\bm \theta, R, Z, L, -M)}{2},
\end{equation}
i.e., the mMTP energy is invariant with respect to inversion of magnetic moments. Thus, mMTP explicitly depends on magnetic moments.

Each simulation using mMTPs consists of two stages. At the first stage, we minimize the energy with respect to magnetic moments, i.e., we equilibrate magnetic moments for fixed atomic positions and lattice vectors. At the second stage, we move the atoms and change the lattice vectors in accordance with a concrete atomistic simulation (e.g., relaxation, molecular dynamics (MD), etc.) for fixed magnetic moments. Thus, the simulation at finite temperature used here differs from spin-lattice dynamics, where the spin values evolve according to physical equations \cite{landau1935theory, gilbert1955lagrangian}. Therefore, we conclude that magnetic moments play an important role in mMTPs and are considered as an additional degree of freedom alongside with atomic coordinates, atomic types, and lattice vectors. For this reason, active learning algorithms for an automated constructing a training set become even more important for magnetic potentials than for non-magnetic ones. 

\subsection{Active learning of magnetic Moment Tensor Potentials}

Manual construction of training sets using trial-and-error method requires significant computational resources. 
In order to automate this process, we proposed an active learning (AL) procedure for selecting the most informative configurations during an atomistic simulation with a pre-trained magnetic MTP. 
Afterwards, we conduct DFT calculations for the selected configurations to update the training set and re-fit the potential.
To that end, we generalized the active learning algorithm for non-magnetic MTPs from \cite{podryabinkin2017-AL} and \cite{gubaev2018machine} to magnetic MTPs.

To construct the initial training set including $K$ configurations with energies $E_k^{\rm DFT}$, forces $\textbf{f}_{i,k}^{\rm DFT}$, stresses $\sigma_{ab,k}^{\rm DFT}$, and magnetic forces (the negative sign of the derivative of energy with respect to the magnetic moment is used by convention because we consider how the magnetic moment of an atom affects other atoms by changing their energy) $T_{i,k}^{\rm DFT}$, $k = 1, \ldots, K$ we utilized Density Functional Theory (DFT) and constrained DFT (see the following section).
To fit the initial mMTP, we minimize the following objective function:

\begin{equation}
\label{eq:loss_func}
\begin{split}
    \sum_{k=1}^{K} \Biggl[ w_e \Bigl(E^{\rm mMTP}_k(\bm \theta) - E_k^{\rm DFT} \Bigr)^2 \\ + w_f \sum_{i=1}^{N} \Bigl(\textbf{f}_{i,k}^{\rm mMTP}(\bm \theta) - \textbf{f}_{i,k}^{\rm DFT} \Bigr)^2 \\ + w_s \sum_{a, b = 1}^{3} \Bigl(\sigma_{ab,k}^{\rm mMTP}(\bm \theta) - \sigma_{ab,k}^{\rm DFT} \Bigr)^2  \\ + w_t \sum_{i=1}^{N} \Bigl(T^{\rm mMTP}_{i,k}(\bm \theta) - T^{\rm DFT}_{i,k} \Bigr)^2 \Biggr]
\end{split}
\end{equation}
with respect to the parameters $\bm {\theta}$, where $w_e, w_f, w_s$, and $w_t$ are non-negative weights. This objective function \eqref{eq:loss_func} was introduced in \cite{Kotykhov2024_mag_force_fit} for mMTPs where it was shown that fitting to magnetic forces improves the reliability of mMTPs.

To formulate an active learning algorithm, we start from \eqref{eq:loss_func}. Assume, we have obtained the vector of optimal mMTP parameters $\bm {\bar{\theta}}$ of length $m$ after minimizing the functional \eqref{eq:loss_func} with $w_e=1, w_f=0, w_s=0$, and $w_t=0$, i.e., we parameterized mMTP only on DFT energies. We then linearize the first term in \eqref{eq:loss_func}:
$$E^{\rm DFT}_k - E^{\rm mMTP}_k({\bm {\theta}}) \approx E^{\rm DFT}_k - \sum \limits_{p=1}^{m} (\theta_p - \bar{\theta}_p) \dfrac{\partial E^{\rm mMTP}_k({\bm {\bar{\theta}}})}{\partial \theta_p}.$$
We can then interpret the fitting of mMTPs as the solution of the following overdetermined system of equations with respect to $\theta_p$:
\begin{align} \label{eq:OverdeterminedSystem}
\sum \limits_{p=1}^{m} \theta_p \dfrac{\partial E^{\rm mMTP}_k({\bm {\bar{\theta}}})}{\partial \theta_p} = E^{\rm DFT}_k + \sum \limits_{p=1}^{m} \bar{\theta}_p \dfrac{\partial E^{\rm mMTP}_k({\bm {\bar{\theta}}})}{\partial \theta_p}.
\end{align}
The matrix of the system \eqref{eq:OverdeterminedSystem} is:
\begin{equation}
\label{eq:MatrixB}
    B = \begin{pmatrix}
\frac{\partial E^{\rm mMTP}}{\partial \theta_1} (\bm {\bar{\theta}}, {\bm x}^{(1)}) & \dots & \frac{\partial E^{\rm mMTP}}{\partial \theta_m} (\bm {\bar{\theta}}, {\bm x}^{(1)}) \\
\vdots & \ddots & \vdots \\
\frac{\partial E^{\rm mMTP}}{\partial \theta_1} (\bm {\bar{\theta}}, {\bm x}^{(K)}) & \dots & \frac{\partial E^{\rm mMTP}}{\partial \theta_m} (\bm {\bar{\theta}}, {\bm x}^{(K)})
\end{pmatrix},
\end{equation} 
where $\bm{x}^{(1)}, \ldots, \bm{x}^{(K)}$ are the atomic configurations in the training set corresponding to the rows of the matrix \eqref{eq:MatrixB}. From the $K \times m$ matrix $B$ we select an $m \times m$ submatrix $A$ that maximizes absolute value of the determinant (volume), thus, we select a set of the $m$ most linearly independent rows in the matrix $B$.
This is the core idea of our active learning algorithm, which is based on the D-optimality criterion.

For constructing the matrix $A$ with the $m$ most linearly independent rows of $B$, we use the maxvol algorithm \cite{goreinov2010_maxvol}, which ensures diversity of the configurations corresponding to the rows of $A$. If $K \geq m$, we select $m$ configurations out of the $K$ configurations for which the absolute value of the volume $|{\rm det}(A)|$ is maximal. If $K < m$, we select some (or even all) of the $K$ configurations which give the maximal $|{\rm det}(A)|$ and fill in the rest of the rows with ones on the diagonal and with zeros elsewhere.
As defined above, each configuration $\bm{x}$ depends on a set of lattice vectors $L$, on a set of atomic positions $R$, on a set of atomic types $Z$, and on a set of magnetic moments $M$. Therefore, the matrix $A$ also depends on magnetic moments and this enables us to select configurations with fixed magnetic moments.

\begin{figure*}[!ht]
\begin{center}
\includegraphics[scale=0.5]{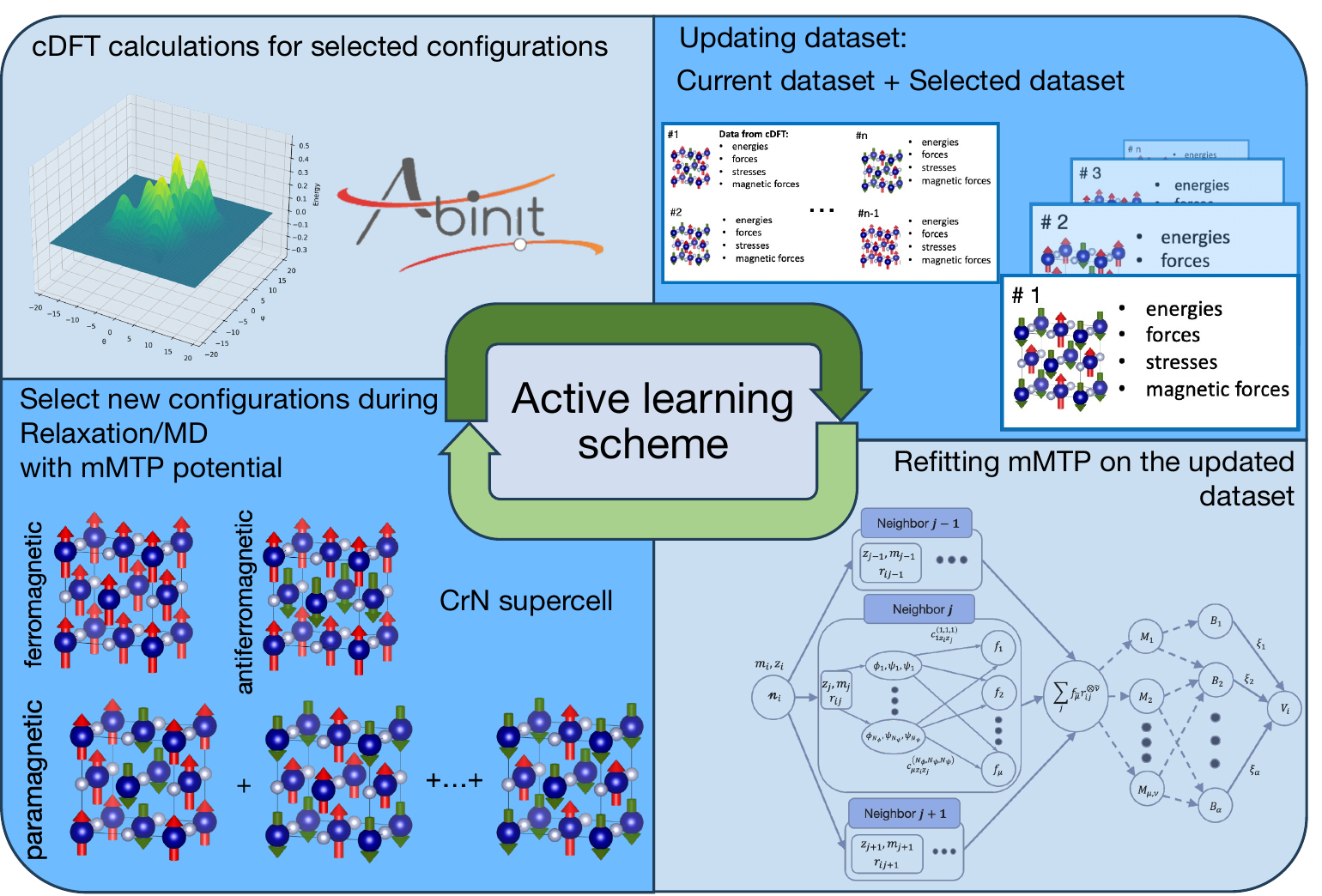}\caption{Protocol of magnetic MTP active training. We start from an initial mMTP (the scheme of mMTP is also shown in Fig. \ref{fig:mMTP}) and an initial training set. Next, we conduct active learning during magnetic moment equilibration, relaxation, or MD simulation. If the extrapolation grade of any configuration is too high then we terminate a simulation and select new configurations from the ones obtained during this simulation (the preselected configurations). After that, we carry out cDFT calculations for the selected configurations. Finally, we update the training set and re-fit the mMTP. We re-start the process from the very beginning until no configuration is preselected during equilibration/relaxation/MD simulation.}
\label{fig:TOC}
\end{center}
\end{figure*}

After training the initial mMTP and constructing the matrix $A$, we start an atomistic simulation and compute the extrapolation grade for each configuration ${\bm x^*}$ encountered during this simulation:

\begin{equation}
\label{eq:extrapol_grade}
    \gamma({\bm x^*}) = \max\limits_{1 \le j \le m} |c_j|,
\end{equation}
where the vector $\textbf{c} = (c_1, \ldots, c_m)$ for the configuration ${\bm x^*}$ is computed by:
\begin{equation}
\label{eq:c}
    \textbf{c} = \Bigl(\frac{\partial E^{\rm mMTP}}{\partial \theta_1} (\bm {\bar{\theta}}, {\bm x}^*), \dots, \frac{\partial E^{\rm mMTP}}{\partial \theta_m} (\bm {\bar{\theta}}, {\bm x}^*) \Bigr) A^{-1}.
\end{equation}
The extrapolation grade $\gamma$ is the amount by which $|{\rm det}(A)|$ increases if ${\bm x^*}$ is added to the training set. During a simulation, we create a set of preselected candidate configurations potentially to be added to the training set.
To that end, we introduce two thresholds: $\gamma_{\rm low}$ and $\gamma_{\rm up}$. The first threshold, $\gamma_{\rm low}$, is a lower bound of permissible (or, reliable) extrapolation, and the second threshold, $\gamma_{\rm up}$, is an upper bound of permissible extrapolation. If $\gamma_{\rm low} \leq \gamma({\bm x^*}) \leq \gamma_{\rm up}$ then we add the configuration ${\bm x^*}$ to the preselected set. In this case, the extrapolation is considered as reliable, i.e., the errors in predicting energies, forces, stresses, and magnetic forces, by the mMTP are still acceptable to continue the atomistic simulation. If $\gamma({\bm x^*}) > \gamma_{\rm up}$, we consider the extrapolation as too risky.
We then terminate the atomistic simulation, add this configuration to the preselected set, and update the matrix $A$ with the configurations from the preselected set using the maxvol algorithm, i.e., we select the new configurations to be added to the training set. We now conduct the DFT calculations for the selected configurations, update the training set, and re-fit the mMTP. We then restart the atomistic simulation, select new configurations to be added to the training set, and re-fit the mMTP again. We continue this process until no configurations are added to training set during the entire simulation. The above AL protocol is shown in Fig. \ref{fig:TOC}. 

In the previous section, we emphasized that one of the stages of modeling using mMTP involves the equilibration of magnetic moments. Thus, the configurations with non-equilibrium magnetic moments occur during an atomistic simulation. Therefore, these configurations can be selected using our AL algorithm. Calculating the energies, forces, stresses, and magnetic forces of these configurations requires constrained DFT (cDFT) calculations. Constrained DFT enables for imposing hard constraints on the magnetic moments.

\subsection{Constrained Density Functional Theory calculations}

In conventional spin-polarized DFT, we minimize the Hohenberg-Kohn total energy functional $E[\rho; R, Z, L]$ with respect to the electron density $\rho = \rho(\mathbf{r})$, where $R$ denotes nuclear coordinates (or, atomic positions), $Z$ denotes atomic types, $L$ is a set of lattice vectors:
\begin{equation} \label{eq:electron_density}
\begin{aligned}
\rho({\bf r}) &= \sum_{i=1}^{N_e} \int \psi_i^{\uparrow *} ({\bf r}) \psi_i^{\uparrow} ({\bf r}) \, d^3 {\bf r} + \\
&\quad \sum_{i=1}^{N_e} \int \psi_i^{\downarrow *} ({\bf r}) \psi_i^{\downarrow} ({\bf r}) \, d^3 {\bf r} = {\rho}^{\uparrow}({\bf r}) + {\rho}^{\downarrow}({\bf r}),
\end{aligned}
\end{equation}
where \( \psi_i^{\uparrow} ({\bf r}) \) and \( \psi_i^{\downarrow} ({\bf r}) \) are the electron wavefunctions corresponding to the spin up and spin down electrons, respectively. By solving the following minimization problem:

\begin{equation}
\label{eq:DFT}
E_{\rm DFT}(R, Z, L) = \min_{\rho} E[\rho; R, Z, L],
\end{equation}
we obtain the optimal electron density $\rho^* = {\rho}^{\uparrow *} + {\rho}^{\downarrow *}$. We note that positions of nuclei $R$, lattice vectors $L$, and atomic types $Z$ are fixed when solving equation \eqref{eq:DFT}. Using the optimal electron densities ${\rho}^{\uparrow *}$ and ${\rho}^{\downarrow *}$, the magnetic moment of the $i$-th atom can be calculated by integrating over a specific region $\Omega_i$ around the atom:

\begin{equation} \label{magnetic_moment_equil}
    m_i^* = \int_{\Omega_i} \left( {\rho}^{\uparrow *}({\bf r}) - {\rho}^{\downarrow *}({\bf r}) \right) d {\bf r}.
\end{equation}
However, in the case of conventional spin-polarized DFT calculations, the resulting magnetic moments $m_i^*$, $i = 1, \ldots, A, \ldots, N$ depend on the ground-state spin up and spin down electron densities. Therefore, magnetic moments cannot be considered as an additional fixed degree of freedom along with atomic positions, atomic types, and lattice vectors.

In this work, we used the constrained density functional theory (cDFT) method proposed by Gonze \textit{et al.} in \cite{Gonze_2022}. This method is specifically designed to achieve the same self-consistent solution as determined by the relevant governing equations. The cDFT energy functional is defined as follows:

\begin{equation} \label{EcDFT}
E^{\rm cDFT}_{v_{\rm ext},m_{\rm A}}[u] = E^{\rm v}_{v_{\rm ext}}[u] - R^{\rm v}_{\rm A}[u] (W_{\rm AA})^{-1} \left(m^{\rm v}_{\rm A}[u] - m_{\rm A}\right).
\end{equation}

Here, $E^{\rm v}_{v_{\rm ext}}[u]$ is the  DFT energy, $W_{\rm AA}$ is the integral of a weight function $w_{\rm A}(\mathbf{r})$ (which is 1 inside the volume $\Omega_{\rm A}$ associated with fragment A, and 0 outside) squared, $u$ represents the screened potential, $v_{\rm ext}$ is the external potential depending on atomic positions and lattice vectors, $m^{\rm v}_{\rm A}[u]$ is the magnetic moment described with Eq.~\ref{magnetic_moment_equil}, but with spin up and spin down electron densities, $\rho^{\uparrow}$ and $\rho^{\downarrow}$, self-consistently potential solved for the functional \eqref{EcDFT}, and $m_{\rm A}$ is the fixed magnetic moment for a specific atomic fragment A.
The residual self-consistent potential $R^{\rm v}_{\rm A}[u]$ accounts for discrepancies between the input and output screening potentials within the designated region. Importantly, the functional $E^{\rm cDFT}_{v_{\rm ext},m_{\rm A}}[u]$ is stationary when evaluated at the self-consistent potential. This means that in the self-consistent solution, the energy functional reaches an equilibrium state where its variation with respect to the screened potential $u$ is zero.

In other words, in this cDFT method, atomic positions and lattice vectors enter the screened potential and are treated on the same footing with $m_{\rm A}$, namely, as external parameters of the calculation. Thus, we minimize the constrained energy functional $E^{\rm cDFT}_{v_{\rm ext},m_{\rm A}}[u]$, enabling the spin up and spin down electron densities, $\rho^{\uparrow}$ and $\rho^{\downarrow}$, to vary only under the constraint that $m_{\rm A}$ remains fixed during each self-consistent iteration.

Thus, minimization of $E^{\rm cDFT}_{v_{\rm ext},m_{\rm A}}[u]$ enables us to calculate the configurations selected during any atomistic simulation using the AL algorithm at the fixed atomic positions, atomic types, lattice vectors, and magnetic moments. Therefore, cDFT constitutes the crucial component of the AL scheme illustrated in Fig. \ref{fig:TOC}.   

\subsection{Computational details}

For DFT and constrained DFT (cDFT) calculations \cite{Gonze_2022}, the ABINIT package \cite{gonze2020abinit, romero2020abinit} was used with an energy cutoff of 680 eV and a $4\times4\times4$ k-point mesh.
For describing the exchange-correlation functional in DFT, we used the local-density approximation (LDA) with the Hubbard on-site correction $U$ of 3 eV. This value was chosen for CrN in \cite{alling2010effect, zhou2014structural} based on a comparison of the lattice parameter of B1-CrN in the paramagnetic state, and the valence band electronic density of states of cubic B1-CrN in the paramagnetic state, with experimental data.

We used an mMTP of level 16 with $N_{\phi} = 8$ and $N_{\psi} = 2$ for the two-component CrN system which results in 606 parameters ${\bm \theta}$.
The minimum distance between the atoms $R_{\rm min}$ and the cut-off radius $R_{\rm cut}$ were set to 1.7 \AA ~and 5.0 \AA, respectively.
The values of 3.2 $\mu_B$ for the Cr atoms and 0.1 $\mu_B$ for the N atoms were chosen as the maximum magnetic moments $M_{\rm max}^{\rm Cr}$ and $M_{\rm max}^{\rm N}$.
We also choose the weights of the objective function \eqref{eq:loss_func} to be $w_{\rm e} = 1$, $w_{\rm f} = 0.01$ ~\AA$^2$, $w_{\rm s} = 0.001$, and $w_{\rm t} = 0.1 ~\mu^{2}_{B}$.
Active learning was performed with $\gamma_{\rm low} = 2$ (during relaxation), $\gamma_{\rm low} = 4$ (during molecular dynamics), and $\gamma_{\rm up} = 10$. The choice of $\gamma_{\rm low}$ significantly affects the number of configurations selected during active learning and the final accuracy of the potential. As demonstrated in the paper \cite{podryabinkin2017-AL}, the recommended value of $\gamma_{\rm low}$ lies between 2 and 11. An increase in this value results in a decrease in the number of configurations selected. For active learning during molecular dynamics, we used a larger $\gamma_{\rm low}$ than for relaxation to reduce the number of DFT calculations. Regarding $\gamma_{\rm up}$, it is recommended to use a value not exceeding 20 to maintain the reliability of atomistic simulations (see, e.g. \cite{novikov2020mlip}).
For all the performed calculations, we used an equiatomic configuration of 64 atoms in a supercell.

\section{Results and discussion} \label{ResDiscuss}

\subsection{Training set and fitting errors}

\begin{figure}[!ht]
\begin{center}
\includegraphics[scale=0.20]{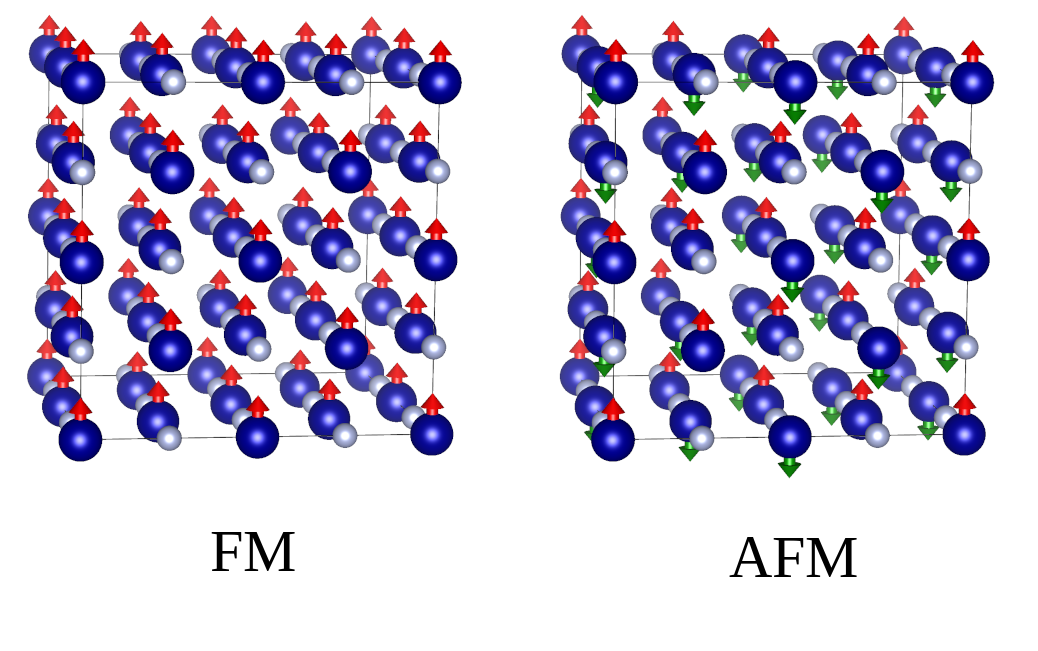}
\caption{B1-CrN in the FM and AFM states. Cr atoms are shown in blue, N atoms are shown in grey.}\label{fig:FM_and_AFM}
\end{center}
\end{figure}

The training set was constructed in several stages. We started from an ideal cubic B1-CrN structure with lattice parameters given by 4.13 \AA, 4.15 \AA, and 4.17 \AA.
The choice of these values is related to the fact that, in \cite{alling2010effect}, the Hubbard on-site correction $U$ was selected in accordance with the experimental data for the lattice parameter of the paramagnetic state at room temperature, which is 4.13 \AA ~according to \cite{corliss1960antiferromagnetic}.
Values greater than this were used to account for thermal expansion at high temperatures. 
For each of these lattice parameters we considered several magnetic states: antiferromagnetic (AFM), ferromagnetic (FM) (see Fig. \ref{fig:FM_and_AFM}), and four randomly disordered collinear magnetic states with 50\% of the magnetic moments oriented upwards and 50\% oriented downwards.
Hence, our initial training set contains 18 configurations.

Next, we took each of these 18 configurations and applied random displacements to each atom.
We repeated this step such that there are now 54 configurations contained in the training set. 
Afterwards, we repeated the previous step for the magnetic moments by applying perturbations, again, two times, by at most 15 \% to each equilibrium magnetic moment from the training set.
As a result, we obtained 108 additional configurations with non-equilibrium magnetic moments that we calculated using cDFT.
The initial mMTP was fitted to this training set containing 162 configurations.

After fitting the initial mMTP, we equilibrated the magnetic moments for each of the configurations in the initial training set.
During the equilibration we used active learning to extract additional extrapolative configurations to be added to the training set;
626 additional configurations were selected during the equilibration.
Next, we conducted full geometrical optimization (relaxation) of the ideal cubic B1-CrN structures in the AFM and FM states with active learning switched-on; 8 more extrapolative configurations were selected and added to the training set.
Afterwards, we constructed 32 configurations by applying shear displacements to the equilibrated configurations of $\pm$1\% along the $xx$ and $yz$ directions for the ferromagnetic and five randomly disordered collinear magnetic states, and along the $xx$, $xy$, $xz$, $zz$ directions for the configurations in the AFM state.
These configurations were calculated with DFT and added to the training set.
Finally, the active learning algorithm was used to select configurations during molecular dynamics (MD) simulations in the NVT ensemble with mMTP at 300 K and 1000 K for FM, AFM, and 10 different randomly disordered collinear magnetic states. The lattice parameters were chosen between 4.12 and 4.20 \AA, with a step of 0.01 \AA. The training set was increased by 760 configurations after the MD simulations at 300 K, and by 835 configurations after MD at 1000 K. We note that various randomly disordered magnetic states were manually created and used as the initial states for the MD simulations. Thus, the signs of the Cr collinear magnetic moments in magnetic states were fixed during MD for all Cr atoms. However, the absolute values of magnetic moments changed during the MD simulations and, thus, we sampled and selected non-equilibrium magnetic configurations with different atomic positions and varying absolute values of magnetic moments during MD.

The final training set contains 2423 configurations. 
We emphasize that about 96\% of the configurations in the training set contain non-equilibrium magnetic moments and were, therefore, calculated with cDFT. 
This highlights the importance of cDFT for fitting mMTP with the AL algorithm. 
The training errors of the mMTP fitted on the final training set can be found in Table \ref{tab:fitting_errors}.
From the table we conclude that the fitting errors are reasonable.

\begin{figure}[!ht]
\begin{center}
\includegraphics[scale=0.15]{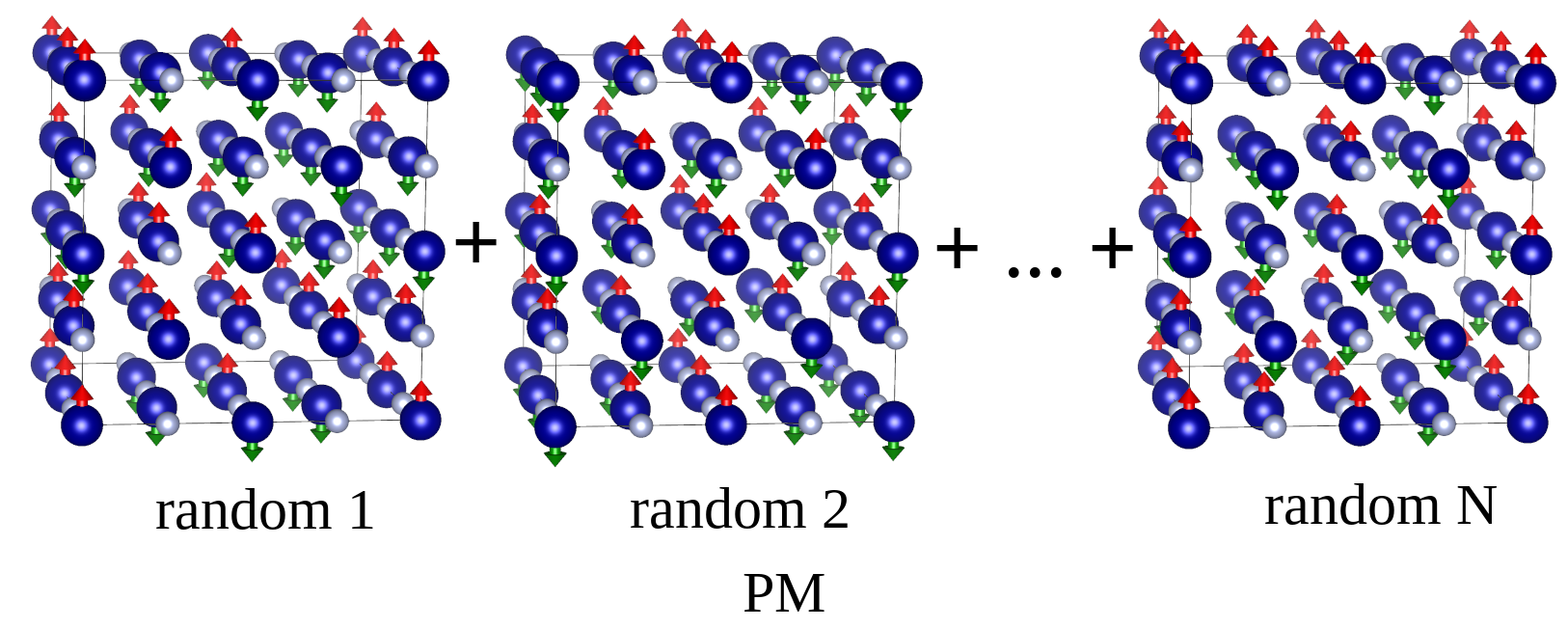}
\caption{B1-CrN in the paramagnetic state is represented as an average over various random disordered collinear magnetic states with 50\% magnetic moments oriented up and 50\% oriented down. Cr atoms are shown in blue, N atoms are shown in grey.}\label{fig:PM}
\end{center}
\end{figure}

\begin{table}[!ht]
\caption{Root mean square errors for energies, forces, stresses, and magnetic forces predicted with the mMTP fitted on the resulting training set.}
\label{tab:fitting_errors}
\begin{center}
\begin{tabular}{cccc}
\hline
energy error  & force error & stress error & magnetic force error\\
(meV/atom) & (meV/\AA) & (GPa) & (meV/$\mu_B$) \\
\hline
1.7 & 108 & 0.4 & 64 \\
\hline
\end{tabular}
\end{center}
\end{table}

\subsection{Averaging over magnetic states}

\begin{figure}[!ht]
    \centering
    \begin{subfigure}{\linewidth}
        \centering
        \includegraphics[width=0.8\textwidth]{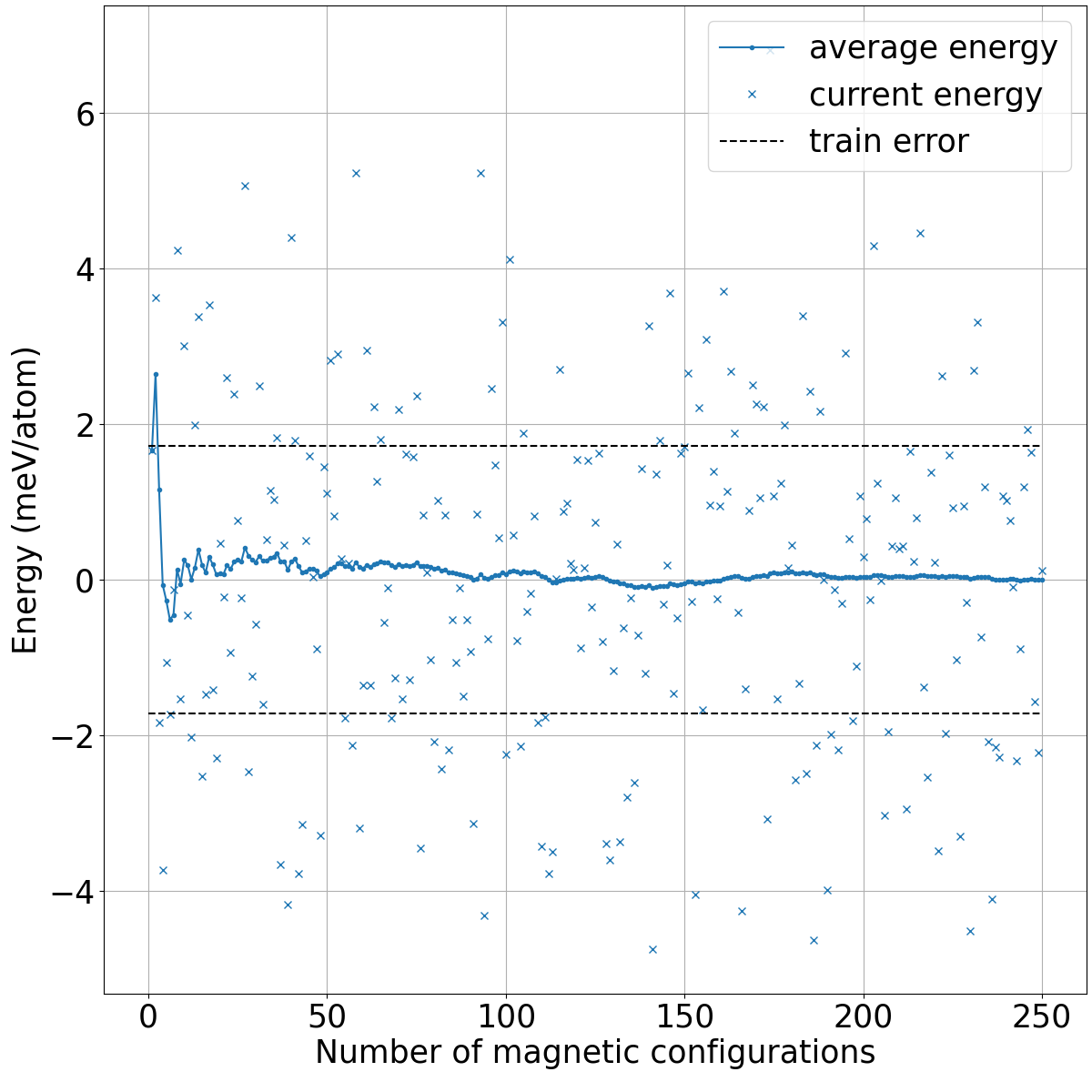}
        \caption{Energy convergence.}
        \label{fig:displaced_atom_energy_convergence}
    \end{subfigure}
    
    \vspace{0.5cm} 
    
    \begin{subfigure}{\linewidth}
        \centering
        \includegraphics[width=0.8\textwidth]{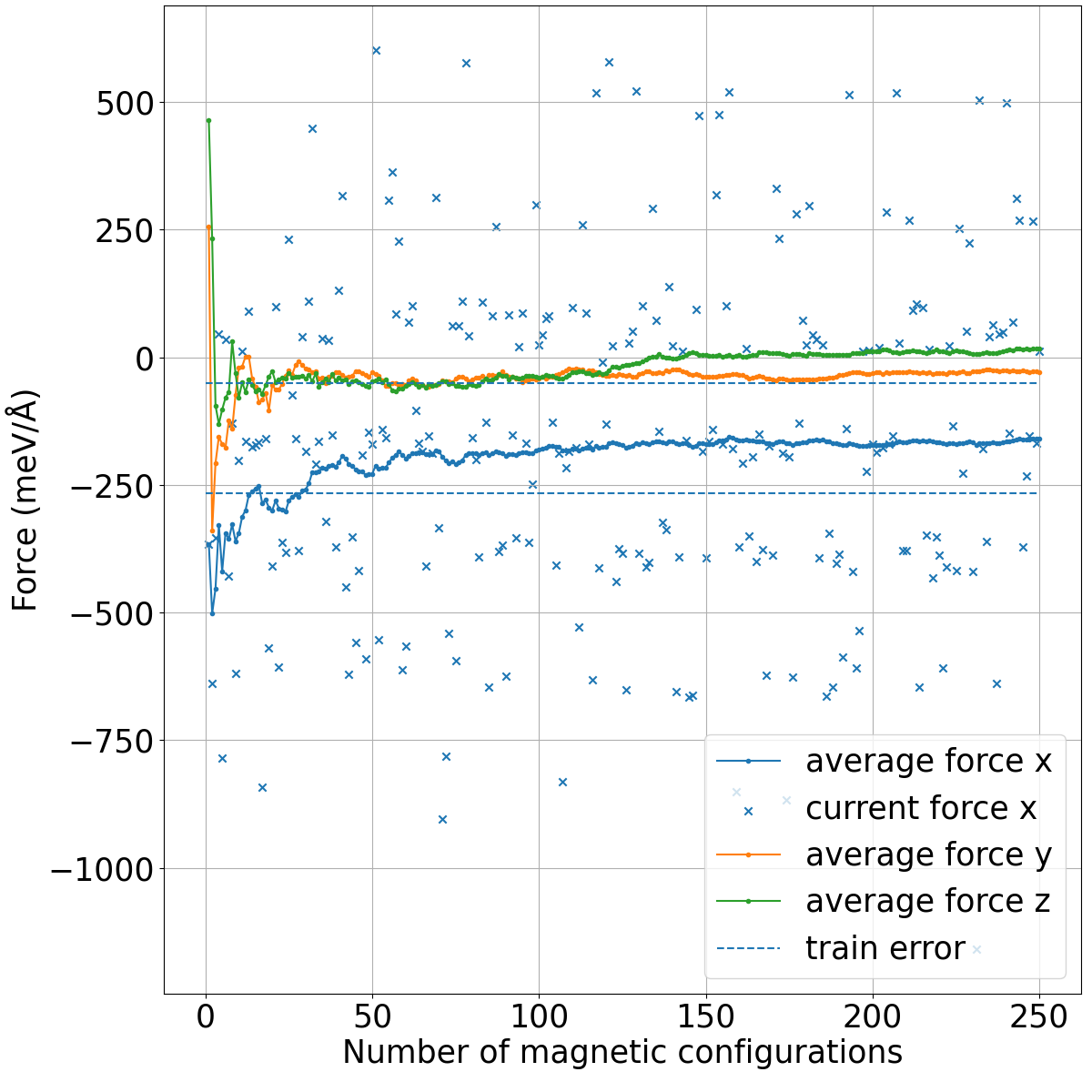}
        \caption{Force convergence.}
        \label{fig:displaced_atom_force_convergence}
    \end{subfigure}

    \caption{Dependence of the averaged forces and energies on the number of magnetic configurations for a system with a shifted Cr atom along the $x$ axis. (\textbf{a}) Points indicate the accumulated average energy at the fixed number of random magnetic configurations, while crosses denote the non-averaged energy of the current magnetic configuration. (\textbf{b}) Points indicate the average force components of displaced Cr atom, and blue crosses denote the non-averaged force along the $x$ axis in the current magnetic configuration.}
    \label{fig:displaced_atom_convergence}
\end{figure}

As previously outlined, the paramagnetic state is obtained by averaging over different randomly disordered collinear magnetic states (see Fig. \ref{fig:PM}).
In \cite{alling2010effect, abrikosov2016recent}, the magnetic sampling method (MSM) was applied to the CrN system in the paramagnetic state to calculate average energy and forces over configurations with randomly disordered magnetic states. In \cite{kormann2012atomic}, the spin-space averaging (SSA) method was proposed to calculate effective atomic forces at finite temperatures by averaging individual forces with Boltzmann weights. In combination with special quasi-random structures (SQS), this method can be used to calculate the phonon spectrum, which was done for CrN in the paramagnetic state \cite{zhou2014structural}.
In this work, we used averaging over different randomly disordered magnetic states to describe the paramagnetic state. Such an averaging corresponds to the limit of infinite temperatures where any magnetic state is equally possible.
Nevertheless, this averaging is suitable for describing the paramagnetic state and predicting the properties of the B1-CrN system (see the results in the subsections below). Initially, the convergence of averaged energies and forces was investigated for a number of magnetic configurations for an ideal crystal with 64 atoms in a supercell, where one atom of Cr was displaced by 0.01 \AA ~along the $x$ axis in Figs. \ref{fig:displaced_atom_energy_convergence} and \ref{fig:displaced_atom_force_convergence}. It can be seen that 50 configurations are already sufficient for averaging, and that the difference between the average forces over the selected number of 250 configurations does not exceed the energy and force fitting error of the mMTP. To further investigate the paramagnetic state, we averaged over 50 randomly disordered magnetic configurations. We note that in \cite{alling2010effect} it was shown that 40 configurations are sufficient to achieve near-zero forces and energy convergence for an ideal crystal of B1-CrN with MSM and, thus, our result is close to the one obtained in this paper. We also note that magnetic SQS and MSM gave almost identical potential energy for this system.

\subsection{Elastic constants}

In the remainder of this section, we discuss the results of the calculations of B1-CrN properties in various magnetic states. We emphasize that, in addition to comparing mMTP to the paramagnetic state observed experimentally, we also compare the mMTP to DFT for the ferromagnetic and antiferromagnetic states in order to validate the developed protocol and the trained mMTP.

First, we compare the elastic constants for various magnetic states. The corresponding results can be found in Table \ref{tab:elastic_constants}---the mMTP and DFT elastic constants coincide very well. From the table we also see that B1-CrN is mechanically stable in all the states as the elastic constants satisfy the Born-Huang stability criterion \cite{born1996dynamical}.

\begin{table}[!ht]
\caption{Elastic constants (in GPa) of B1-CrN in different magnetic states calculated with the fitted mMTP and DFT.}
\label{tab:elastic_constants}
\begin{center}
\begin{tabular}{ccccccccccc}
\hline
Conf. & Method & $C_{11}$ & $C_{22}$ & $C_{33}$ & $C_{12}$ & $C_{13}$ & $C_{23}$ & $C_{44}$ & $C_{55}$ & $C_{66}$ \\
\hline 
\multirow{2}{*}{FM} & mMTP & 594 & 594 & 594 & 100 & 100 & 100 & 188 & 188 & 188 \\
 & DFT[*] & 573 & 573 & 573 & 108 & 108 & 108 & 164 & 164 & 164 \\
 & DFT\cite{zhou2014structural} & 589 & 589 & 589 & 128 & 128 & 128 & 162 & 162 & 162\\
\hline
\multirow{2}{*}{AFM} & mMTP & 620 & 620 & 708 & 113 & 97 & 97 & 158 & 158 & 164 \\
 & DFT[*] & 680 & 680 & 709 & 100 & 62 & 62 & 122 & 122 & 177 \\
 & DFT\cite{zhou2014structural} & 696 & 696 & 722 & 113 & 76 & 76 & 124 & 124 & 174 \\
\hline
\multirow{2}{*}{PM} & mMTP & 645 & 645 & 645 & 98 & 98 & 98 & 160 & 160 & 160 \\
 & DFT\cite{zhou2014structural} & 649 & 649 & 649 & 99 & 99 & 99 & 145 & 145 & 145 \\
\hline
\end{tabular}
\end{center}
\footnotetext{The DFT results indicated by [*] refer to the calculations performed in this work.}
\end{table}

\subsection{Phonon spectra}

\begin{figure}[!ht]
\begin{center}
\includegraphics[scale=0.17]{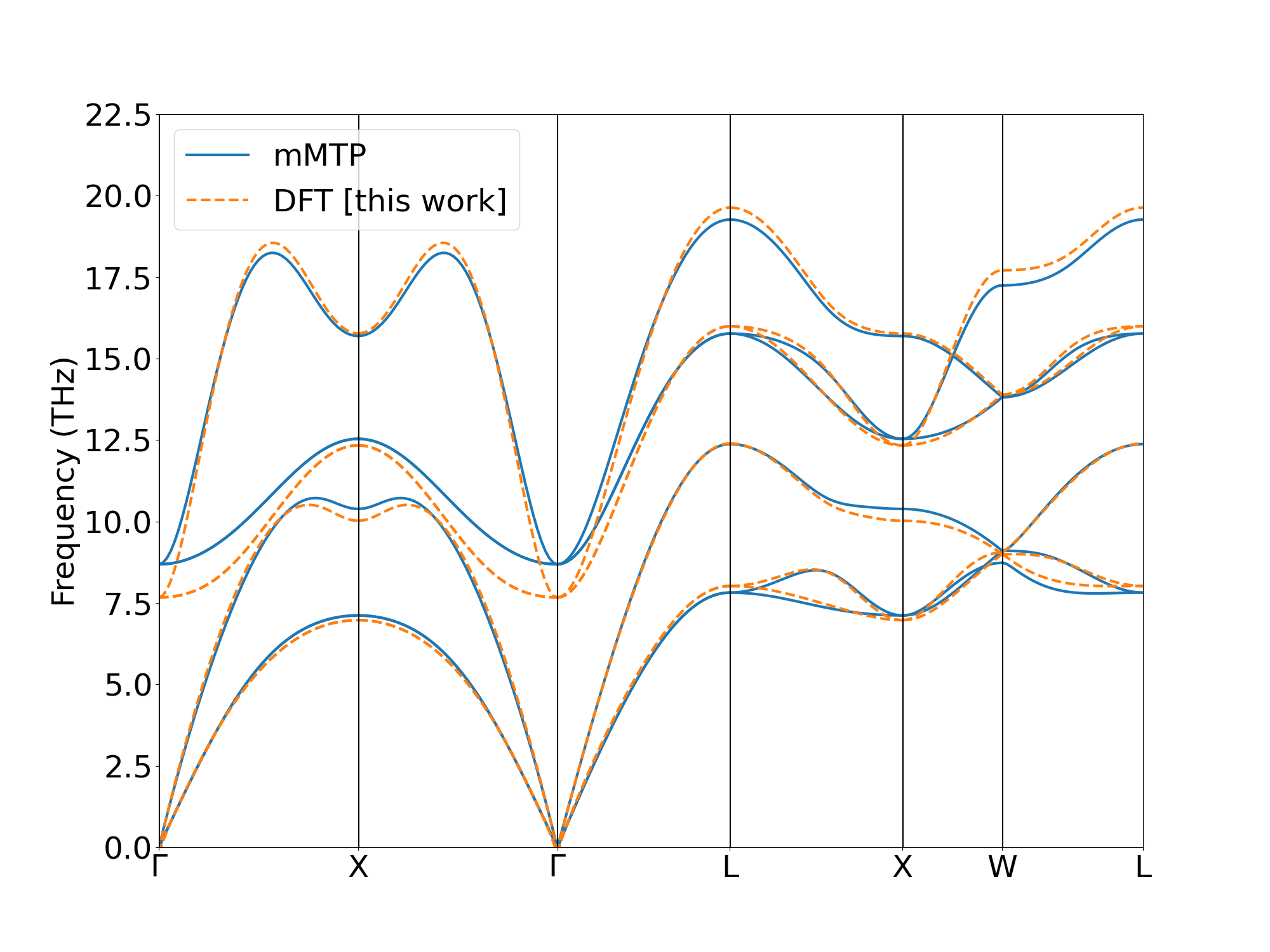}\caption{Phonon spectrum for B1-CrN in the ferromagnetic state calculated with the fitted mMTP (solid lines) and DFT using the ABINIT code (dashed lines).}\label{fig:FM_phonon}
\end{center}
\end{figure}

\begin{figure}[!ht]
\begin{center}
\includegraphics[scale=0.17]{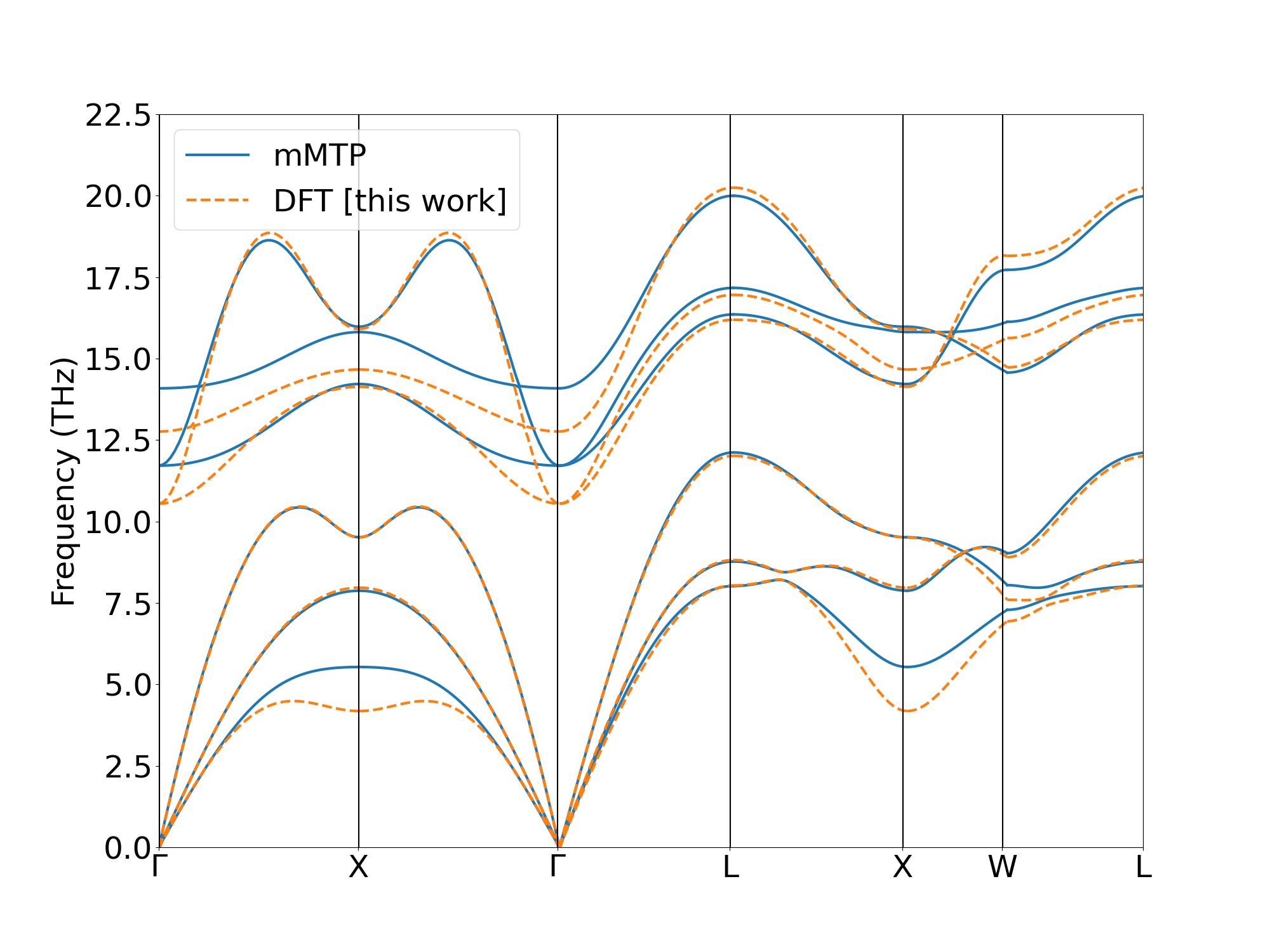}\caption{Phonon spectrum for B1-CrN in the antiferromagnetic state calculated with the fitted mMTP (solid lines) and DFT using the ABINIT code (dashed lines).}\label{fig:AFM_phonon}
\end{center}
\end{figure}

\begin{figure}[!ht]
\begin{center}
\includegraphics[scale=0.17]{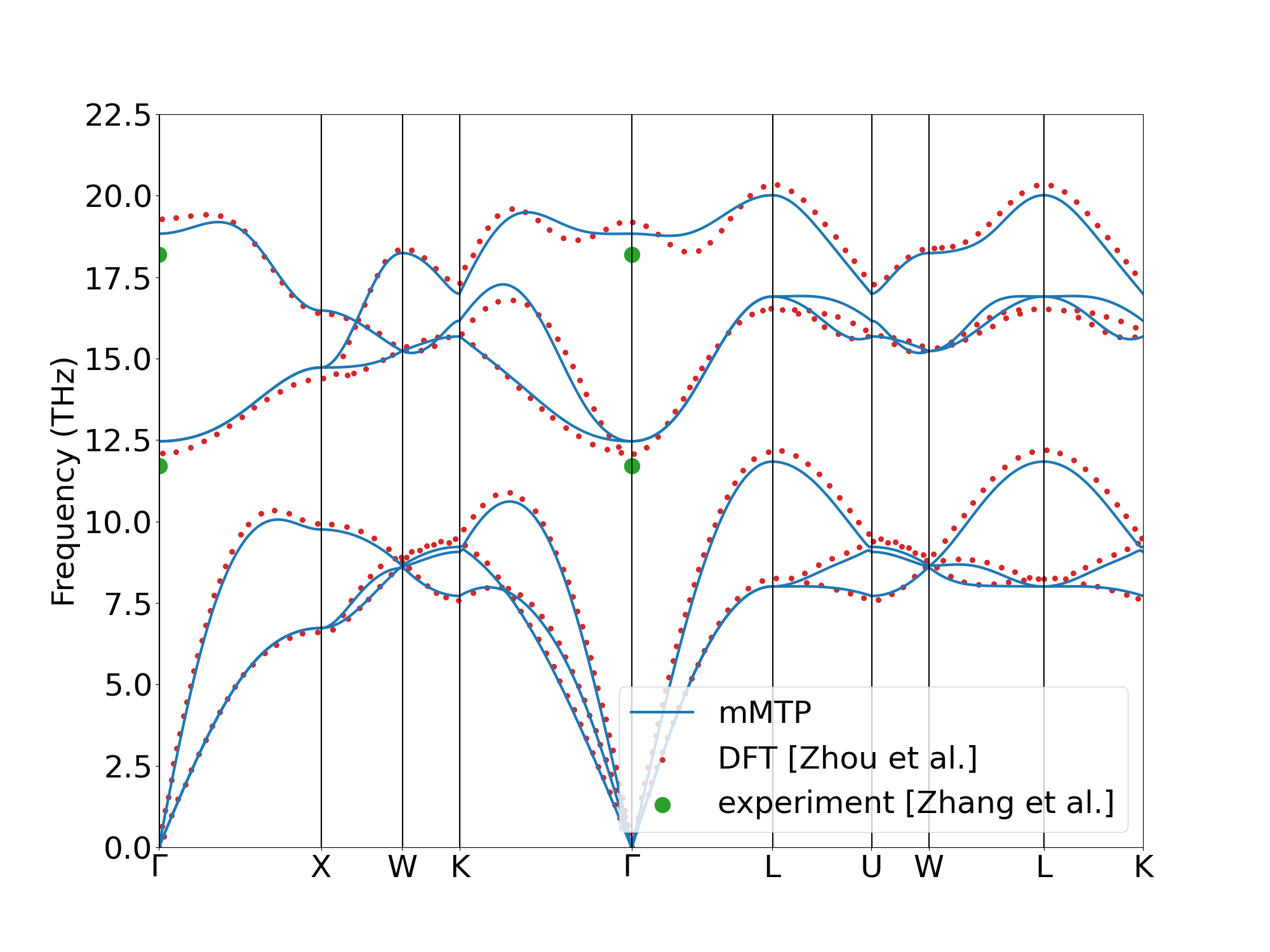}\caption{Phonon spectrum for B1-CrN in the paramagnetic state obtained with the fitted mMTP (solid lines), DFT \cite{zhou2014structural} (points), and experimental values at the $\Gamma$ point from Raman and infrared measurements \cite{zhang2010crn}.}\label{fig:PM_phonon}
\end{center}
\end{figure}

Next, we calculated the phonon spectra for B1-CrN in the ferromagnetic, antiferromagnetic, and paramagnetic states using the PHONOPY package \cite{phonopy-phono3py-JPCM, phonopy-phono3py-JPSJ}. The phonon spectra calculations were conducted with 64-atom 2$\times$2$\times$2 B1-CrN supercells. We compared the mMTP phonon spectra in the paramagnetic state with the DFT phonon spectra reported in \cite{zhou2014structural} and with experimental values at the $\Gamma$ point from Raman and infrared measurements. The phonon spectra are shown in Figs. \ref{fig:FM_phonon}, \ref{fig:AFM_phonon}, and \ref{fig:PM_phonon}, respectively.

For each of the considered magnetic states, the phonon spectra obtained with mMTP and DFT agree very well with each other. We remark that for calculating the phonon spectrum in the paramagnetic state using mMTP, non-analytic term corrections to the dynamical matrix \cite{wang2010mixed} should be applied to describe the longitudinal-transversal optical phonon branches (LO-TO) splitting at the $\Gamma$ point. For the dielectric tensor and the Born effective charge matrix, experimental values were taken from \cite{zhang2010crn}. Due to the symmetry of the system, only the diagonal elements $\varepsilon_{\alpha\alpha}$ of the dielectric tensor are non-zero, and they are equal to each other; their values correspond to 22. Additionally, due to charge neutrality, the diagonal elements of the Born effective charge matrix satisfy $Z = Z_{\alpha\alpha}^{Cr} = -Z_{\alpha\alpha}^{N}$, with a value $Z$ of 4.4. The frequencies calculated near the $\Gamma$ point using mMTP and DFT in \cite{zhang2010crn}, along with corrections to the dynamical matrix, are close to the experimental values.

\subsection{Lattice thermal expansion coefficient and specific heat capacity}

\begin{figure}[!ht]
\begin{center}
\includegraphics[scale=0.3]{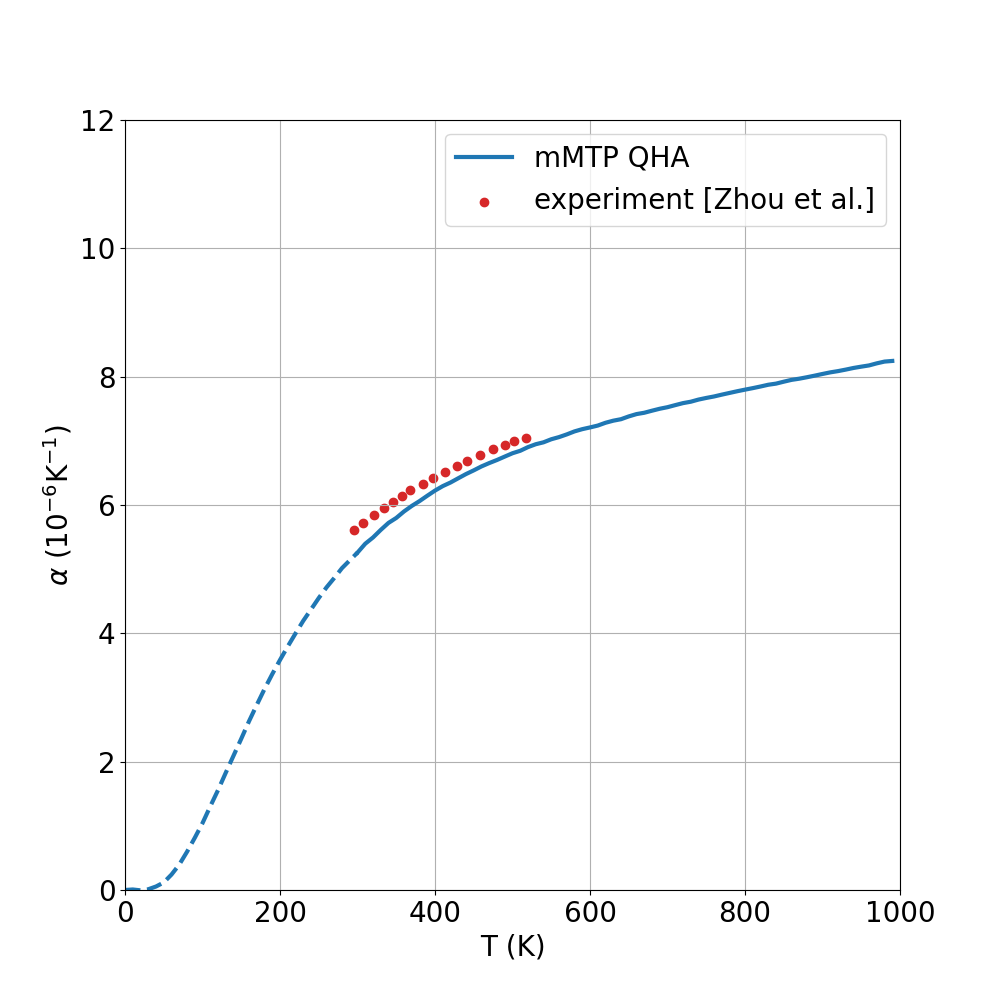}\caption{Linear thermal expansion coefficient for B1-CrN in the paramagnetic state obtained with the fitted mMTP (dashed and solid line) and experimentally \cite{zhou2014structural}. The data below the Neel temperature of 280 K are shown as a dashed line as B1-CrN does not exist below this temperature.}\label{fig:thermal_expansion}
\end{center}
\end{figure}

\begin{figure}[!ht]
\begin{center}
\includegraphics[scale=0.3]{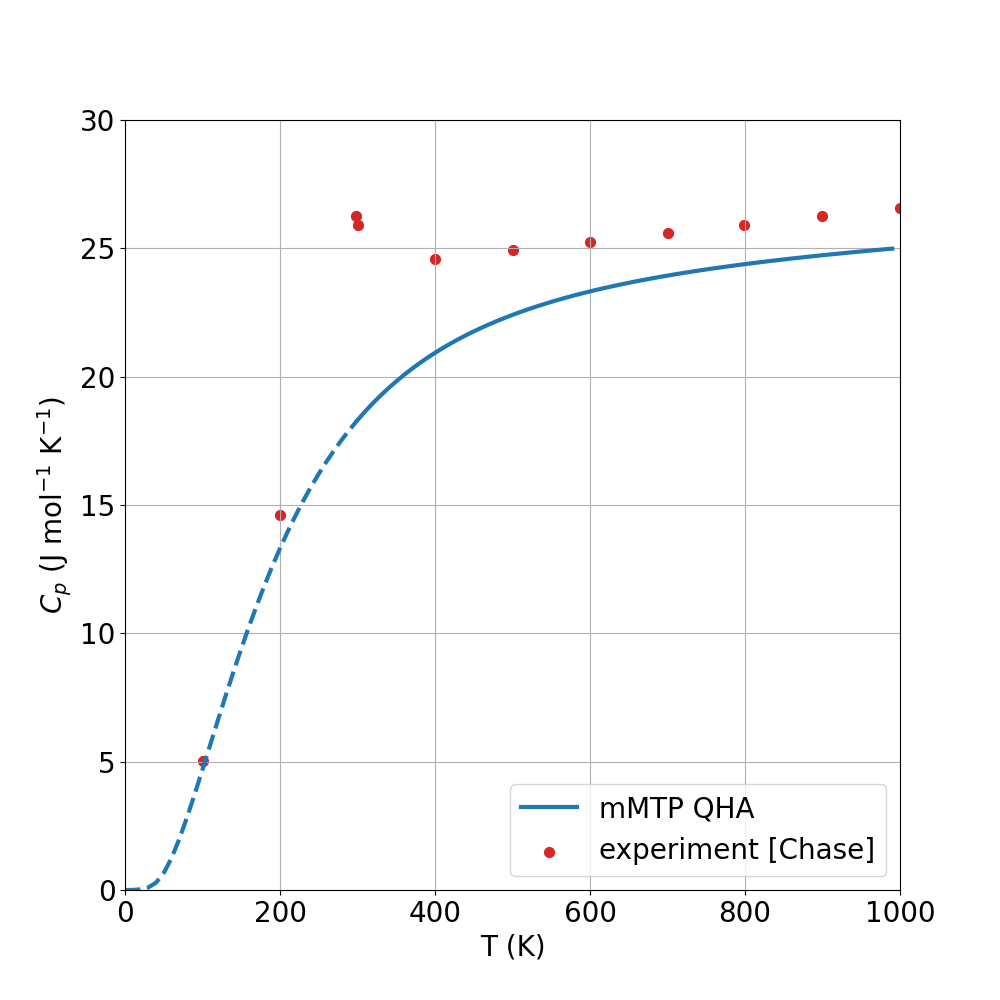}\caption{Specific heat capacity for B1-CrN in the paramagnetic state calculated with the fitted mMTP (dashed and solid line) and experimentally \cite{chase1998nist}. The data below the Neel temperature of 280 K are shown as a dashed line as B1-CrN does not exist below this temperature.}\label{fig:heat_capacity}
\end{center}
\end{figure}

In \cite{zhou2014structural}, it was shown that the anharmonic effects are negligibly small in CrN: the maximum deviation between the vibrational free energies calculated with and without anharmonic effects was about 5 meV/atom. Therefore, the quasi-harmonic approximation (QHA) can be used for predicting the lattice thermal expansion and the specific heat capacity. We have computed both properties for B1-CrN in the paramagnetic state as shown in Figs. \ref{fig:thermal_expansion} and \ref{fig:heat_capacity}. The calculations were performed with PHONOPY using QHA. The dashed lines indicate that B1-CrN in the paramagnetic state does not exist below the Neel temperature of 280 K (this temperature corresponds to the phase transition from Ortho-CrN in the antiferromagnetic state to B1-CrN in the paramagnetic state) and the solid lines demonstrate that it does exist at the temperatures greater than the Neel temperature.
This fact is known from the experiment \cite{corliss1960antiferromagnetic, browne1970investigation}. From Fig. \ref{fig:thermal_expansion}, it follows that the lattice thermal expansion coefficient obtained with mMTP is in good agreement with the experimental value taken from \cite{zhou2014structural}.

The specific heat capacity calculated with the mMTP corresponds to the experimental value starting from a temperature of 400 K (see Fig. \ref{fig:heat_capacity}). The discrepancy between the mMTP and experimental specific heat capacities between 280 K and 400 K is due to the absence of Ortho‑CrN data in the training set. Since the mMTP was trained solely on B1‑CrN, it lacks the information needed to capture the structural changes associated with the phase transition and the corresponding peak in specific heat capacity. This limitation confines the model's accuracy to environments represented in the training data, and including Ortho‑CrN data would be necessary to improve predictions near and above the phase transition temperature \cite{chase1998nist}. Nevertheless, predicting the phase transition in CrN was not the objective of this work and Figs. \ref{fig:thermal_expansion} and \ref{fig:heat_capacity} demonstrate that the fitted mMTP overall correctly predicts the thermal properties of B1-CrN.

\section{Conclusion} \label{Conclusions}

In this work, we used machine-learning interatomic potential with magnetic degrees of freedom (magnetic Moment Tensor Potential, mMTP) for calculating the properties of B1-CrN in various magnetic states. For the construction of the training set, we have developed an active learning algorithm using density functional theory (DFT) and constrained DFT (cDFT) calculations with hard constraints on magnetic moments which enables us to automate the process of constructing training sets by selecting relevant configurations directly during the equilibration of magnetic moments, optimization of the geometric structure of B1-CrN, or molecular dynamics simulations. The final training set comprises 2423 configurations which is rather small compared to other works on MLIPs for modeling paramagnetism (see, e.g., \cite{Novikov2022-mMTP,drautz2024_noncolACE, yu2024_spinGNN, yu2024_spinGNN++}). We used the actively trained potential to predict elastic constants and phonon spectra in the ferromagnetic, antiferromagnetic, and paramagnetic states and demonstrated that the properties predicted with our mMTP coincide very well with DFT. In addition, we used the quasi-harmonic approximation for predicting thermal properties, namely, the lattice thermal expansion coefficient and the specific heat capacity of paramagnetic B1-CrN. Both thermal properties predicted with the mMTP are in a good agreement with experimental results.

In future, we plan to develop a methodology for conducting active training of mMTP during disordered-local-moment molecular dynamics (DLM-MD) with magnetic moment flips for different materials. This approach will enable us to construct the training set in a more natural way rather than relying on active learning during 10 MD simulations with an ensemble of 10 different randomly disordered collinear magnetic states, as done in the present work. Active learning during DLM-MD will enable us to describe many different magnetic states since random magnetic states will change during DLM-MD (due to the flips of magnetic moments). This guarantees that we obtain a reliable potential without the extrapolation problems in a wide temperature range and for many random magnetic states. Once we have such a reliable potential, we will be able to sample data directly from DLM-MD and to account for, e.g., anharmonic effects in materials where they are of great importance and where the quasi-harmonic approximation is not applicable. In addition, we are planning to develop a methodology for actively training of mMTPs during MD simulations that include Monte Carlo (MC) spin flips (MDMC). As opposed to DLM-MD, MDMC rejects non-physical magnetic states; thus, the randomly disordered magnetic states in the sampled configurations will be related to specific temperatures. This gives us the opportunity to accurately investigate complex magnetic materials.

\section*{Data availability}

The data that support the findings of this article are openly available \cite{data}.

\section*{Acknowledgments}

This work was supported by Russian Science Foundation (grant number 22-73-10206, https://rscf.ru/project/22-73-10206/). We thank Dr. Nikita Rybin for fruitful discussions and Dr. Tatiana Kostiuchenko for artistic illustrations.

Max Hodapp acknowledges the financial support under the scope of the COMET program within the K2 Center “Integrated Computational Material, Process and Product Engineering (IC-MPPE)” (Project No 886385); this program is supported by the Austrian Federal Ministries for Climate Action, Environment, Energy, Mobility, Innovation and Technology (BMK) and for Labour and Economy (BMAW), represented by the Austrian Research Promotion Agency (FFG), and the federal states of Styria, Upper Austria and Tyrol.


%

\end{document}